%% file: main.tex
\documentclass[aip,prl,reprint,nofootinbib]{revtex4-2}

\usepackage{amsmath}
\usepackage{amssymb}
\usepackage{graphics}
\usepackage{graphicx}
\usepackage{psfrag}
\usepackage{subfigure}
\usepackage{color}
\usepackage{url}
\usepackage{enumerate}
\usepackage{rotating}
\usepackage{fancyhdr}
\usepackage[ruled, vlined]{algorithm2e}
\usepackage[symbol]{footmisc}
\usepackage{blindtext}
\usepackage{float}
\usepackage[export]{adjustbox}
\usepackage{bm}
\usepackage{adjustbox}

\usepackage[normalem]{ulem}
\usepackage{color}

\definecolor{gray}{rgb}{0.6,0.6,0.6}

\makeatletter
\def\@dotsep{4.5}
\makeatother

\begin{document}

\title{On the generation and destruction mechanisms of arch vortices in urban fluid flows}

\author{Eneko Lazpita}
\affiliation{\mbox{School of Aerospace Engineering, Universidad Politécnica de Madrid, E-28040, Spain}}
\author{{\'A}lvaro {Mart{\'\i}nez-S{\'a}nchez}}
\affiliation{\mbox{Instituto de Matemática Pura y Aplicada, Universitat Politècnica de València, Camino de Vera,
46024 València, Spain}}
\author{Adrián Corrochano}
\affiliation{\mbox{School of Aerospace Engineering, Universidad Politécnica de Madrid, E-28040, Spain}}
\author{Sergio Hoyas}
\affiliation{\mbox{Instituto de Matemática Pura y Aplicada, Universitat Politècnica de València, Camino de Vera,
46024 València, Spain}}
\author{Soledad Le Clainche}
\affiliation{\mbox{School of Aerospace Engineering, Universidad Politécnica de Madrid, E-28040, Spain}}
\author{R. Vinuesa}
\email{rvinuesa@mech.kth.se}
\affiliation{\mbox{FLOW, Engineering Mechanics, KTH Royal Institute of Technology, SE-100 44 Stockholm,
Sweden}}

\begin{abstract}

Studying and interpreting the different flow patterns present in urban areas is becoming essential since they help develop new approaches to fight climate change through an improved understanding of the dynamics of the pollutants in urban environments. This study uses higher order dynamic mode decomposition (HODMD) to analyze a high-fidelity database of the turbulent flow in various simplified urban environments. The geometry simulated consists of two buildings separated by a certain distance. Three different cases have been studied, corresponding to the three different regimes identified in the bibliography. We recognize the characteristics of the well-known arch vortex forming on the leeward side of the first building and document possible generation and destruction mechanisms of this vortex based on the resulting temporal modes. These so-called vortex-generating and vortex-breaking modes are further analyzed via proper-orthogonal decomposition. We show that the arch vortex plays a prominent role in the dispersion of pollutants in urban environments, where its generation leads to an increase in their concentration; therefore, the reported mechanisms are of extreme importance in the context of urban sustainability.

\end{abstract}

\maketitle
Urban areas are essential in today's society, with 70\% of the world's population expected to live in cities by 2050 \cite{EuropeanCommission}. Furthermore, cities contribute more than 60\% of the total greenhouse gas emissions and are therefore essential players in the fight against climate change\cite{Manoli2019}. Further, sustainable cities is the eleven Sustainable Development Goal of United Nations. Some attempts have been made to implement predictive models and study the physics associated with the turbulent fluid flows in urban areas\cite{Vinuesa2020}. In particular, we want to identify coherent flow features by applying modal decomposition to turbulent urban flows. This would allow a better design of buildings and cities.  

\begin{figure}[H]
    \centering
    \adjustbox{trim={.35\width} {.22\height} {0.32\width} {.01\height},clip}%
  {\includegraphics[width=2.8\linewidth]{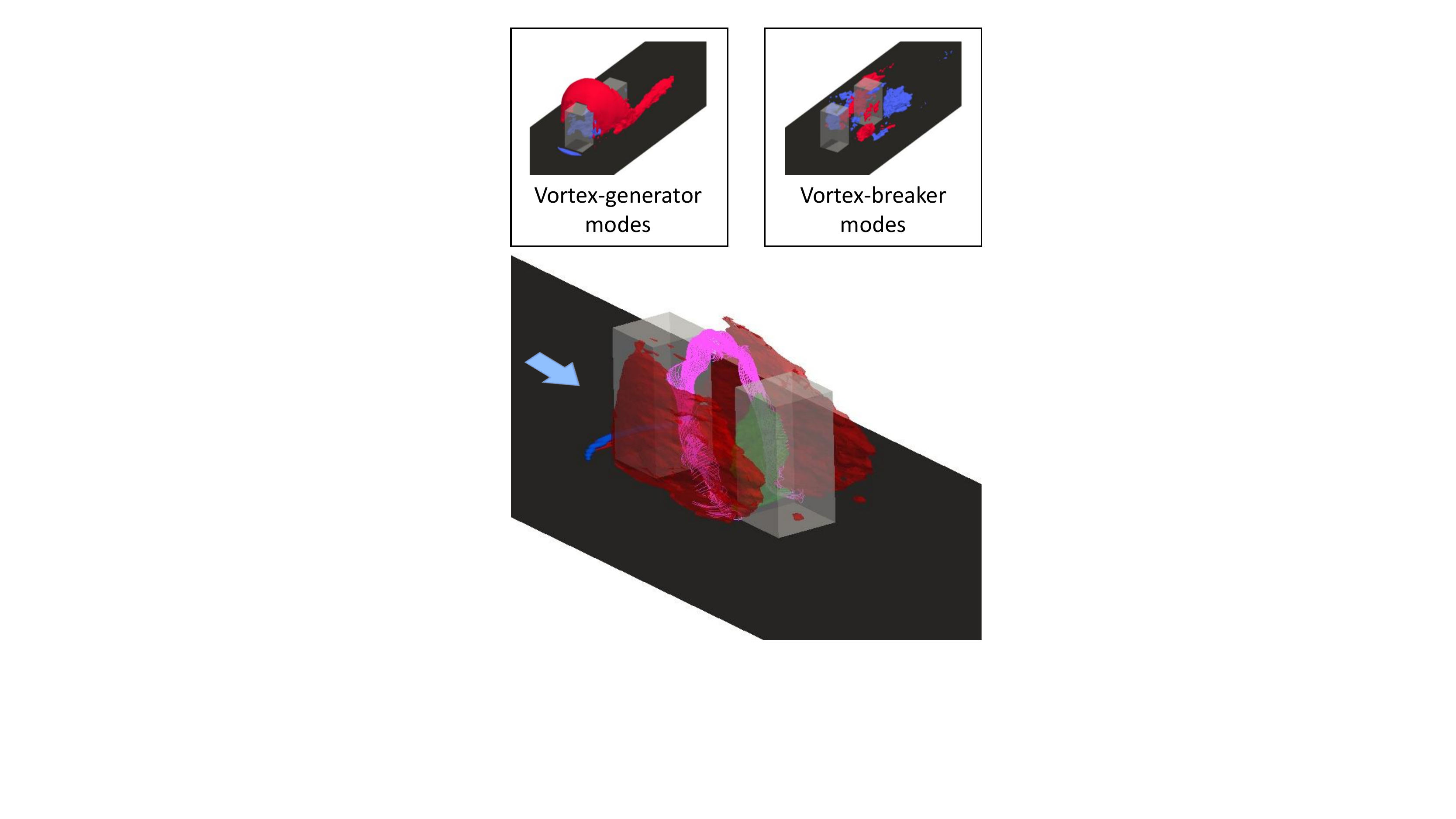}}
	\caption{ Classification of the different modes obtained with HODMD. Bottom pannel: (blue) $v'=0.7$, (red) $u'=0.6$ and (green) $w'=0.8$. Mean streamlines representing the arch vortex are plotted in magenta,  while the two obstacles are in transparent grey. Top panels: streamwise velocity of the modes normalised using the $L_\infty$-norm. The iso-values employed are given by $a\,U_\text{max}$ (red) and $b\,U_\text{min}$ (blue). Top-left: vortex-generator modes with $a=0.8$ and $b=0.5$. Top-right: vortex-breaker modes with $a=0.3$ and $b=0.45$. All cases corresponds to the SF case (see definition below). The arrow indicates the flow direction.   \label{fig:Classification}}
\end{figure}

The arch vortex (which can be observed in Fig.\ref{fig:Classification}) is a vortical structure that appears on the leeward side of a wall-mounted obstacle. It consists of two legs and one roof. In the latter, the flow rotates in the wall-normal direction. In the former, it rotates in the spanwise direction. This vortex has been experimentally studied by Becker et al. (2002)\cite{Becker2002} for different angles of incidence. The wake of the flow around a single obstacle was studied by Kawai et al. (2012)\cite{Kawai2012} using conditional sampling with stereoscopic particle-image velocimetry (SPIV). They obtained an image of the temporal evolution of the arch vortex, which exhibits characteristics similar to those of a K\'arm\'an vortex street. However, the boundary layer used in their study might not have been fully representative of an atmospheric boundary layer (ABL). The coherent structures obtained behind two aligned cubic obstacles were studied by Martinuzzi \& Havel (2000)\cite{Martinuzzi2000} for several distances between the obstacles. AbuOmar \& Martinuzzi\cite{Martinuzzi2003,AbuOmar2008} studied pyramid-shaped obstacles. Simplified urban environments with matrices of obstacles and different angles of incidence (AOI) were studied by Monnier et al. (2018)\cite{Monnier2018}. Other studies have analyzed the flow around obstacles using high-fidelity numerical simulations\cite{Vinuesa2015}.

Another relevant vortical structure is the horseshoe vortex, which is formed on the windward side of the obstacle. It is responsible (together with the vortex-flow pattern) for the flow distribution around an obstacle. The cross-sectional shape of the obstacle has a high influence on the presence and formation of the horseshoe vortex, as discussed by Sumer et al. (1997) \cite{Sumer1997}.  Although both the arch and the horseshoe vortices have been studied in other works mentioned previously, there are several open questions regarding the mechanisms that create and break these structures. In this study we address this critical aspect analyzing the data from a  high-resolution simulation of a urban flow by means of various modal-decomposition algorithms. 

The simulation and modal-decomposition methods are explained in the companion article\cite{PRF}. In both papers, the velocity field is given by $\bm{v}(x,y,z,t)$, where $x, y,$ and $z$ are the streamwise, wall-normal and spanwise directions, respectively, and $t$ is time. Every velocity is normalized with the free stream velocity. The components of the velocity are $\bm{v}=(u,v,w)$, which denote respectively the streamwise, wall-normal and spanwise components. Using Reynolds decomposition, $\bm{v}$ is defined as $\bm{v} = V + \tilde{\bm{v}}$, where $V=\overline{\bm{v}}$ is the average in time and $\bm{\tilde{v}}$ is the turbulent fluctuation. Primes are reserved for intensities $\bm{v'}=\overline{\tilde{\bm{v}}^2}^{1/2}$. 
Briefly, we have simulated the turbulent flow in a simplified urban environment consisting of two obstacles, where different regimes are observed when varying the separation between them\cite{Torres2021}. According to the classification by Oke (1988)\cite{Oke1988}, the first regime is the so-called skimming-flow (SF) configuration, in which the distance between the buildings is the lowest hence the second building hinders the development of the wake of the first one. The second case is denoted as wake-interference (WI) regime, and in this configuration the effect of the second building on the wake of the first one is less pronounced than in the SF case, but still significant. The last case is the isolated-roughness (IR) regime, which is characterized by the highest separation between buildings, and exhibits two different phenomena: the first building behaves as an isolated obstacle and the second building essentially has no effect on the development of the wake from the first one. In this work we focus in the WI case, as the main structures can be clearly observed. Further information about the structures that appear in the rest of the cases are shown in the companion article\cite{PRF}.

Two modal-decomposition methods has been used in this work:proper-orthogonal decomposition {(POD)\cite{POD}} and higher-order dynamic-mode decomposition (HODMD)\cite{Clainche2020}. The HODMD decomposes spatio-temporal data $\bm{v}(x,y,z,t_{k})$, collected at time instant $t_k$ (for convenience expressed as $\bm{v}_{k}$), as an expansion of $M$ modes $\bm{u}_m$, which are weighted by an amplitude $a_m$ as:

\begin{equation}
 \bm{v}(x,y,z,t_{k})\simeq  
 \sum_{m=1}^M a_{m}\bm{u}_m(x,y,z)e^{(\delta_m+i \omega_m)t_k},\label{ab00}
\end{equation}
for $k=1,\ldots,K$.  These modes oscillate in time with frequency $\omega_m$ and may grow, decay or remain neutral in time according to their growth rate $\delta_m$. 

The dominant mode (i.e. the one with the highest amplitude) is located between the frequencies $\omega_m=1$ and $\omega_m=1.2$, while the rest of the modes are subharmonics and harmonics of it. In addition, another relevant mode is the one with the lowest frequency (with a value of $\omega_m=0.11$) since it is the first mode to appear in the spectrum and the periodicity of the main physics is led by its frequency.  

\begin{figure}[H]
    \centering
	\begin{subfigure}
	    \centering
	    \includegraphics[width = 0.49\textwidth]{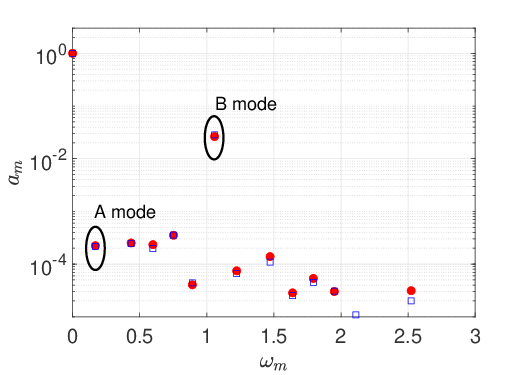}
	\end{subfigure}
	\caption{Spectrum showing the amplitudes versus the frequencies of the temporal modes obtained with HODMD for the WI case. Some A and B modes are highlighted in this figure.  \label{fig:SpectrumHODMD}}
\end{figure}

Based on Fig. \ref{fig:SpectrumHODMD} we identify two types of modes, which are shown in Fig. \ref{fig:Classification}: the vortex-generator (A) modes and the vortex-breaker (B) modes. The vortex-generator modes, as their name indicates, are the ones that generate the main structures and vortices. Therefore, they are related to the mechanism that could create the arch and horseshoe vortex. The A modes usually appear in the low-frequency area of the spectrum and have lower amplitude than the B modes. Some of those modes are $\omega_m=0.12$, $\omega_m=0.24$ or $\omega_m=0.33$. The vortex-breaker modes are the ones that break the main structures of the flow and create the turbulent wake. Unlike the vortex-generators, the B modes appear at high frequencies, from the eighth harmonic of the lowest frequency onward. Some breaker-modes are characterized by the following frequencies: $\omega_m=1.10$, $\omega_m=1.22$ or $\omega_m=1.38$.
\begin{figure*}
    \centering
	\includegraphics[width=0.275\textwidth]{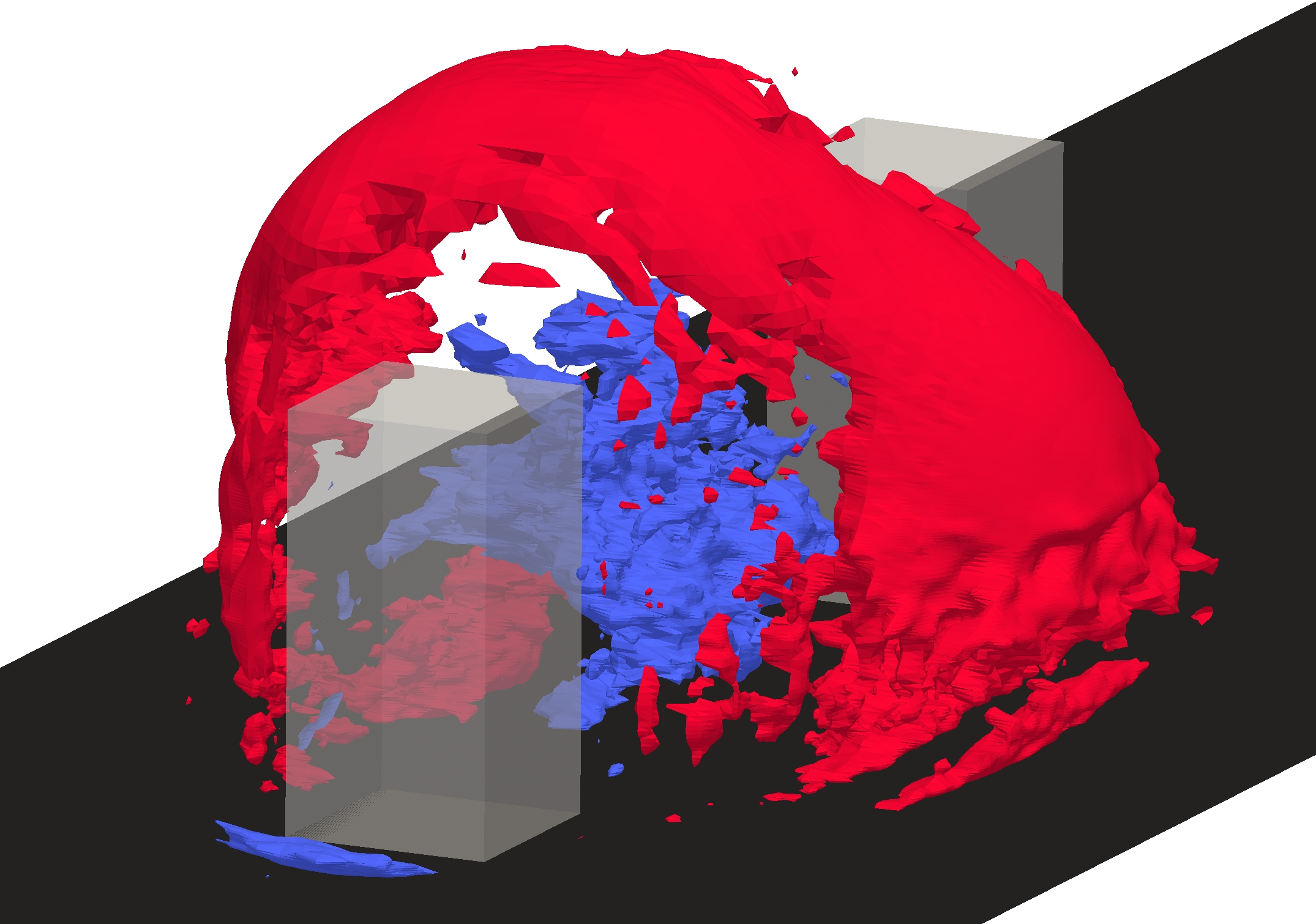}
	\hspace{0.025\textwidth}
	\includegraphics[width=0.275\textwidth]{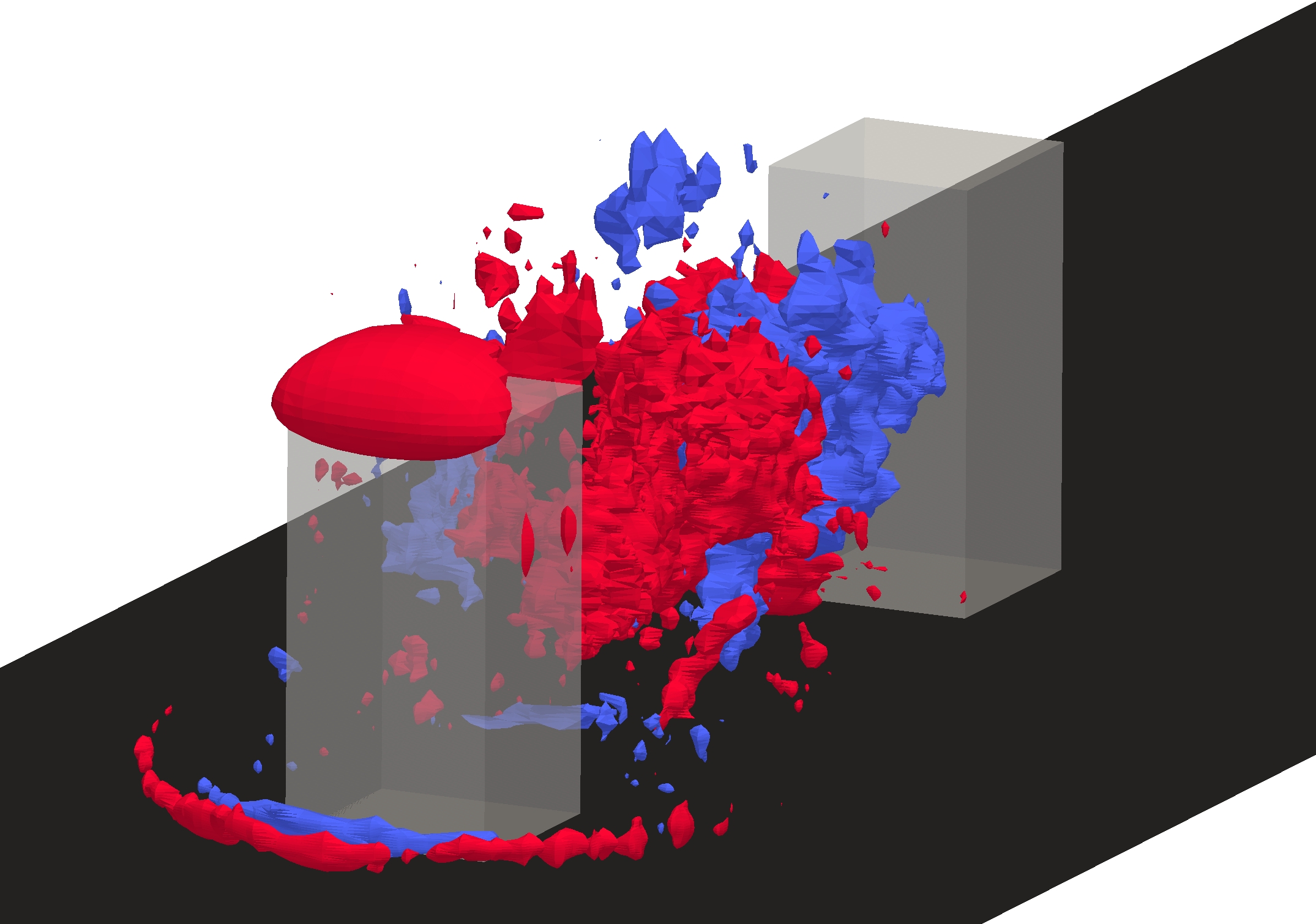}
	\hspace{0.025\textwidth}
	\includegraphics[width=0.275\textwidth]{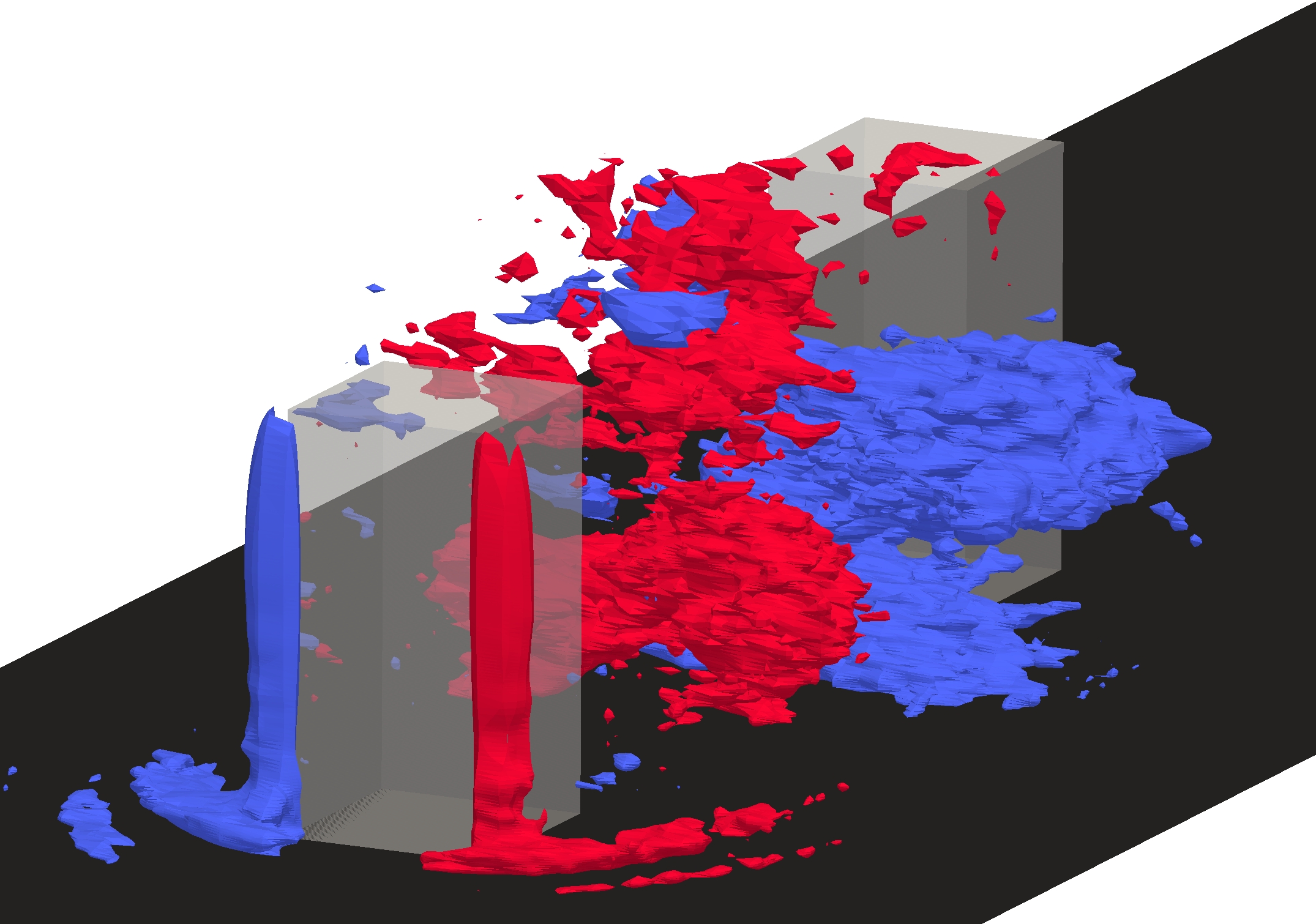}
	\caption{Isosurfaces of the A mode with $\omega_m=0.11$. Velocity values of this mode are normalised using the $L_\infty$-norm. The iso-values employed are given by $a\,U_\text{max}$ (red) and $b\,U_\text{min}$ (blue): Left: Streamwise velocity with $a=0.3$ and $b=0.6$; Middle: wall-normal velocity with $a=0.3$ and $b=0.4$; Right: spanwise velocity with $a=0.45$ and $b=0.45$.\label{fig:Temporal_A_3D}}
\end{figure*}

\begin{figure*}
    \centering
	\includegraphics[width=0.275\textwidth]{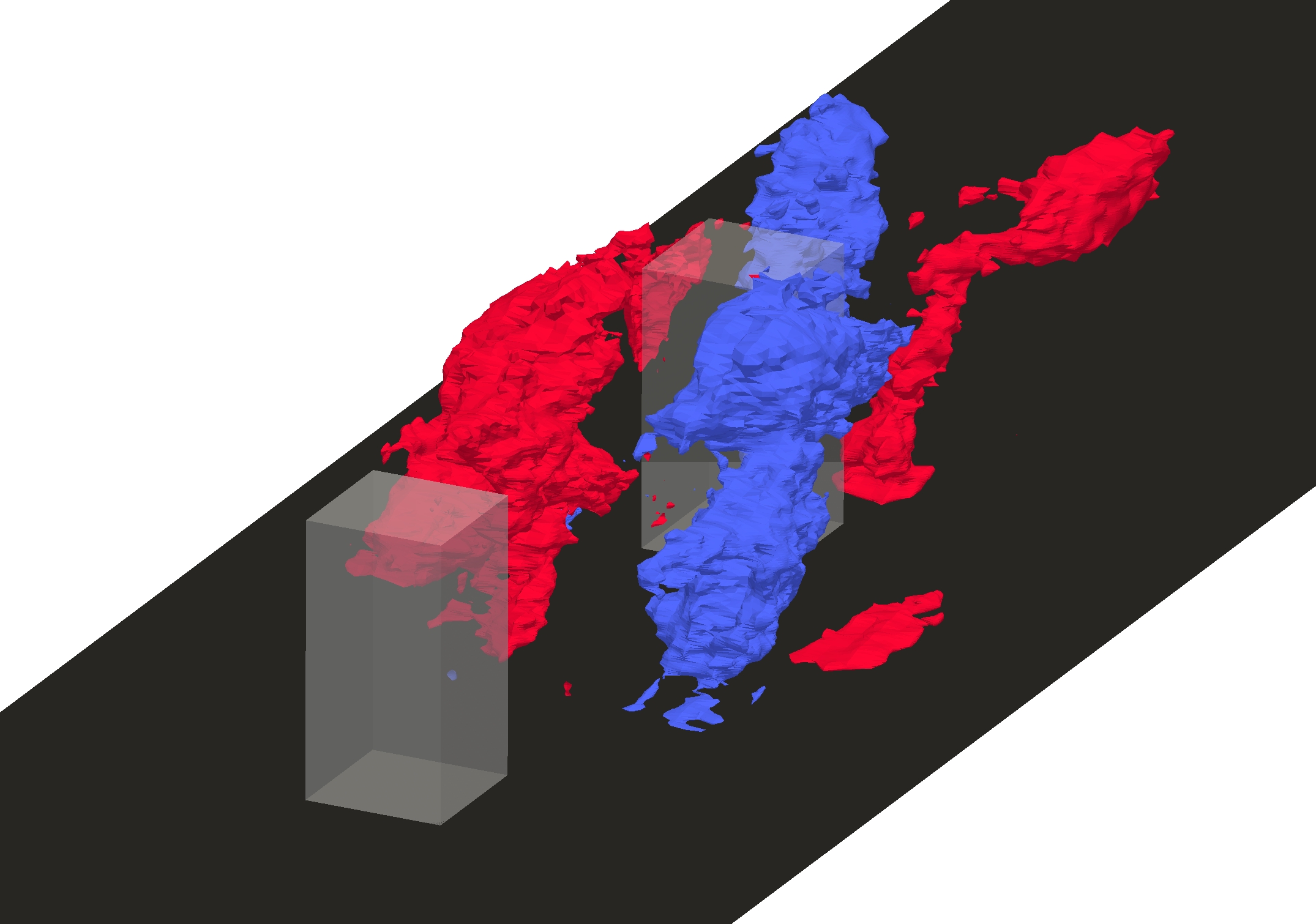}
	\hspace{0.025\textwidth}
	\includegraphics[width=0.275\textwidth]{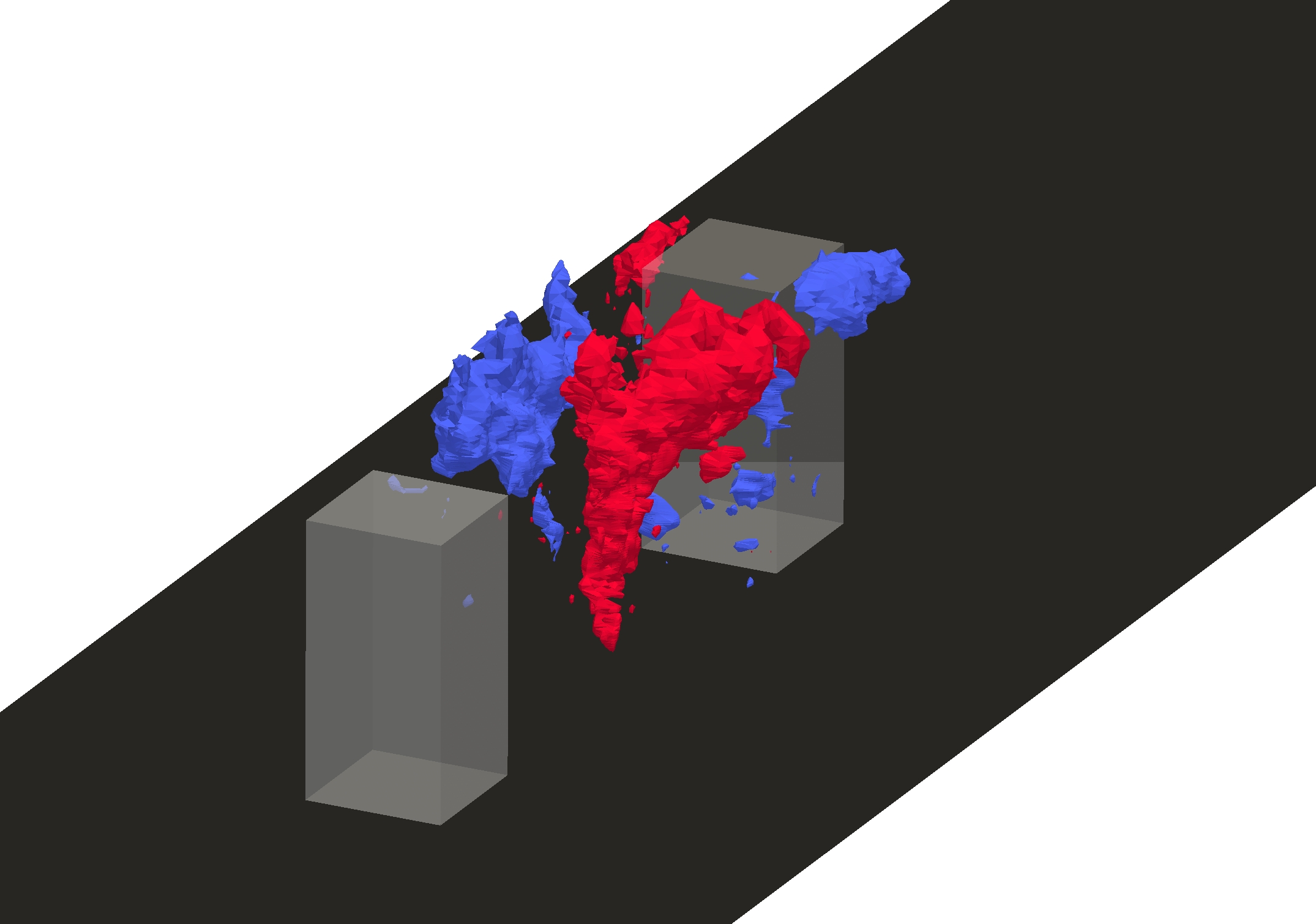}
	\hspace{0.025\textwidth}
	\includegraphics[width=0.275\textwidth]{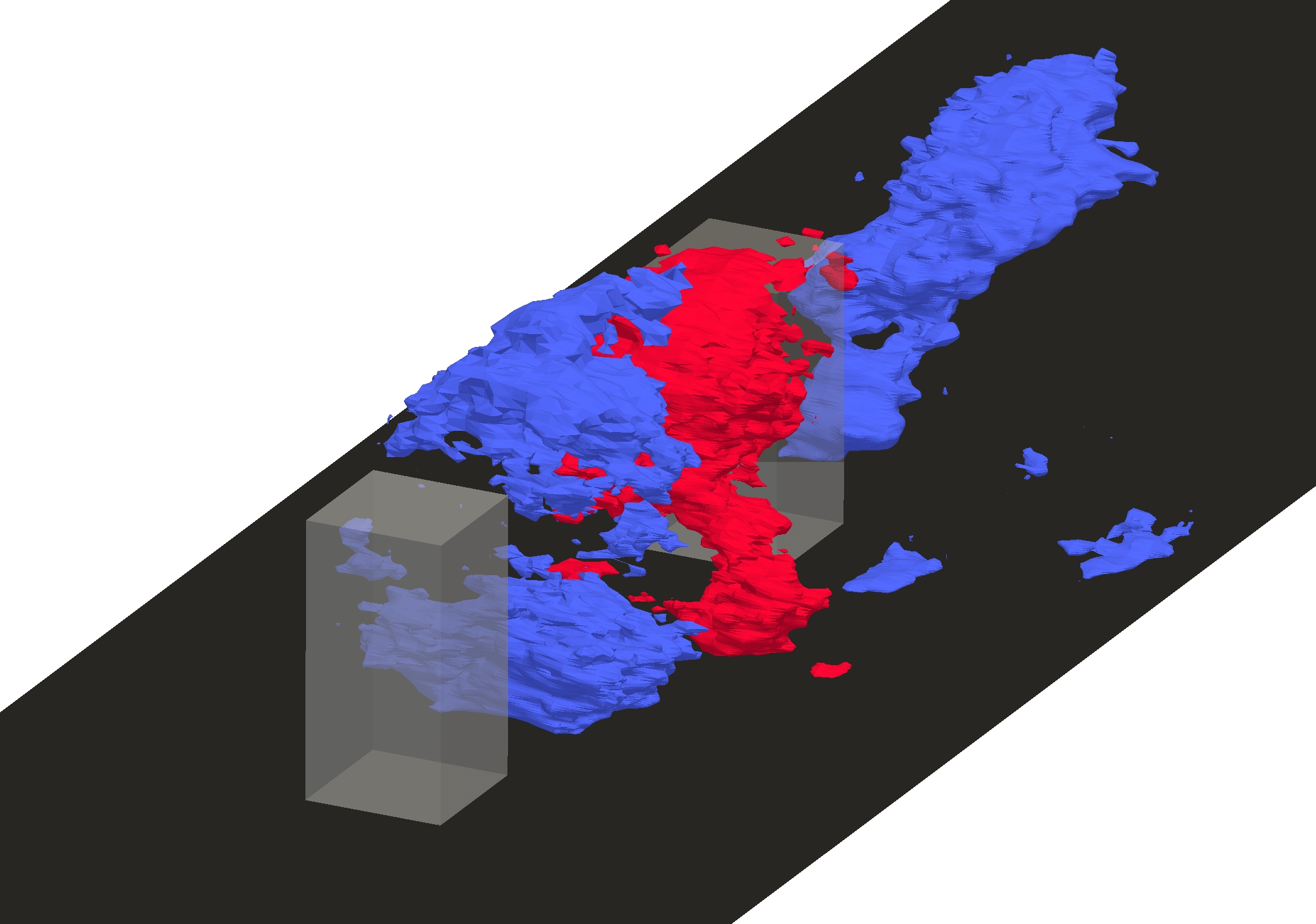}
	\caption{Isosurfaces of the B mode with $\omega_m=1.1$. Velocity values of this mode are normalised using the $L_\infty$-norm. The iso-values employed are given by $a\,U_\text{max}$ (red) and $b\,U_\text{min}$ (blue): (left) Streamwise velocity with $a=0.3$ and $b=0.5$; (middle) wall-normal velocity with $a=0.5$ and $b=0.4$; (right) spanwise velocity with $a=0.4$ and $b=0.5$. \label{fig:Temporal_B_3D}}
\end{figure*}
It has already been shown in the literature that regions of strong recirculation lead to increased concentration of passive scalars\cite{Zhu2017}. As the B modes are related to breaking the coherent structures present in the turbulent flow, they could be connected with the reduced pollutant concentration, while the A modes could be related to high pollutant concentration\cite{Monnier2018}. More specifically, the arch vortex is formed by a flow-recirculating area between the two obstacles. This organization suggests that controlling the presence of the A modes may provide insight into strategies to reduce pollution in urban areas. Further research should be carried out to identify A and B modes in databases modelling multi-phase flows, but this remains an open topic for future research.

The mode with the lowest frequency,  $\omega_m=0.11$, has been selected as the most representative A mode. It is connected with the vortex generation mechanism. As shown in Fig. \ref{fig:Temporal_A_3D}, three main mechanisms have been identified to construct the distinct vortices: a dome, a cap, and two columns for the streamwise, wall-normal, and spanwise velocities, respectively. The dome is a streamwise structure located just after the first building. It surrounds and interacts with the arch vortex by limiting its expansion and recirculating the flow inside the vortex. The cap-like structure is created on top of the first building, interacting primarily with the roof of the arch vortex. This velocity component also exhibits a structure that resembles the horseshoe vortex at the base of the first building. The most representative structure appearing in the spanwise velocity is a column on each side of the first building, affecting the arch vortex legs by creating an ascending rotation of the surrounding flow.

Regarding the influence of the generation mechanisms on the pollutant dispersion, the most unfavourable features are the dome and the cap since they could be connected with a delay in the pollutant escape vertically. The columns appearing buildings side in the spanwise direction have a lower influence. They prevent the flow from extending away from the sides of the buildings but not from escaping from the urban area.

\begin{figure*}
    \centering
    \adjincludegraphics[width=\textwidth,trim={{.025\width} {.035\width} {.025\width} 0}]{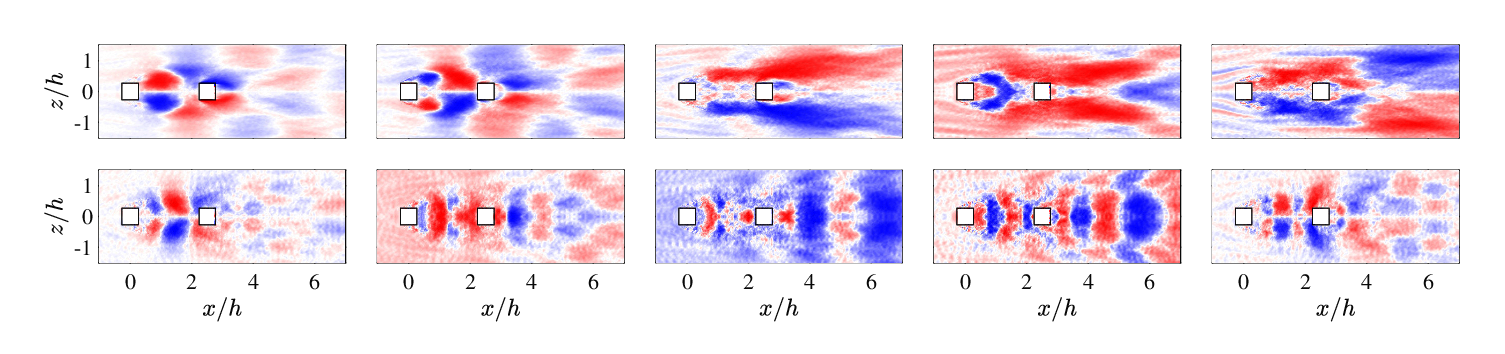}
	\adjincludegraphics[width=\textwidth,trim={{.025\width} {.035\width} {.025\width} 0}]{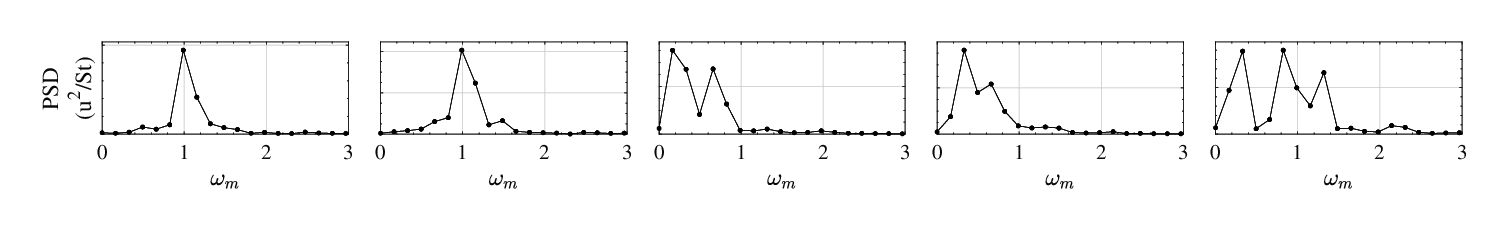}
	\caption{(From left to right) First five POD modes at $y/h=0.25$ (where $h$ is the obstacle height) for the WI case, where we show (top) the streamwise and (middle) spanwise components. Contours are normalised with the $L_\infty$-norm and range between $-1$ (blue) and $+1$ (red). (Bottom) Power-spectral density scaled with the Strouhal number $St=f L/U_{\infty}$ of the temporal coefficients from the POD modes, where $f$ is the characteristic frequency of each mode, $L=h$ is the characteristic length and $U_{\infty}$ is the freestream velocity.   \label{fig: POD}}
\end{figure*}

As in the A modes, the dominant vortex-breaker mode, i.e., the one with the highest amplitude of the whole spectrum, is shown in Fig. \ref{fig:Temporal_B_3D} to assess its main structures. This mode has a frequency of approximately $\omega_m=1.1$. In this type of mode, three additional structures can be distinguished, one for each velocity component: turbulent wake, coherent cluster between the buildings, and an arrowhead-like shape. The streamwise turbulent wake consists of high-velocity coherent clusters on both sides of the buildings. For the wall-normal direction, the high-velocity clusters appear within the width of the buildings. The arrowhead-like shapes that can be seen in the spanwise direction align with the buildings and are inclined downstream. There is a high recirculation area between the structures with positive and negative fluctuations in the B modes, which is related to the creation of a tunnel-shaped vortical structure between the buildings when breaking the arch vortex.

For the destruction mechanisms, the tendency of the streamwise and wall-normal mechanisms is to break the structures by creating clusters that make the flow escape through the sides of the buildings, while the spanwise arrowhead-like structures may help the flow (and the pollutants with it) to escape from the city in a more direct way. 

Furthermore, this differentiation between vortex-generating and -breaking modes can be also examined using POD modes, with which those energetically-important features can be extracted from the flow. Based on the singular-value decomposition (SVD), the POD method decomposes a set of instantaneous fields or snapshots into a set of spatial and temporal modes and their associated singular values\cite{POD}. Here, the aim is to analyse the spatial and temporal modes for the wake-interference regime and state the similarities with the aforementioned modes. Further, details on the POD technique and its results are provided in the companion article\cite{PRF}. 

In Fig. \ref{fig: POD}, the spatial modes for the streamwise and spanwise velocity fields are depicted. Besides, an analysis of the temporal coefficients associated with these modes is performed in the frequency domain through the fast Fourier transform {(FFT)\cite{FFT}} method. This modal-decomposition technique clearly differentiates between low- and high-frequency phenomena, the features of which are very relevant to the vortex-generating and -breaking processes, respectively. Indeed, the first two modes, with associated frequency matching that of the HODMD B modes ($\omega_{m}=1$), are characterised by high-velocity streamwise fluctuations on both sides of the building, which are complemented by spanwise velocity fluctuations aligned with the obstacles. The interaction of these two types of structures results in a tunnel-shaped structure, which suggests that is responsible for the breaking process of the main time-averaged flow structures \cite{PRF}. Therefore, the vortex-breaking process has been identified as the most energetically-relevant mode present in the flow field.

The vortex-generating modes, i.e. the A modes, are related with the third and fourth POD modes, due to their low-frequency behaviour ($\omega_{m}=0.16$ and $\omega_{m}=0.32$). Here, the streamwise component shows how a dome-like structure encloses the region in between the obstacles and it further develops in the wake. This behaviour is similar to that of the time-averaged field, thus inducing a generating process of such structures. Finally, the fifth mode can be regarded as an interaction of the above-mentioned low- and high-frequency modes, which yields flow structures resulting from the combination of such modes. 

To sum up, it has been shown that the turbulent flow in urban environments can be analysed by means of different modal-decomposition algorithms. The HODMD algorithm has shown some DMD modes could be connected to the main mechanisms to generate and break the dominant structures, such as the arch or the horseshoe vortices. Therefore, the obtained temporal modes have been classified into two different groups depending on what mechanisms they contain, the vortex-generator modes or the vortex-breaker modes.

The former appear in the low-frequency region of the spectrum (e.g. $\omega_m=0.11$) and they generate the vortical structures through three different mechanisms: the dome in the streamwise direction, the cap in the wall-normal direction and two columns in the spanwise direction. Meanwhile, the latter appear in the high-frequency portion of the spectrum (e.g. $\omega_m=1.1$) and break the main structures creating the turbulent wake via three mechanisms: the high-velocity clusters on the sides of the buildings in the streamwise direction, other high-velocity clusters that appear between the buildings for the wall-normal velocity and, lastly, the arrowhead-like structures that appear between the buildings in the spanwise direction. 

As mentioned above, the coherent structures may delay the dispersion of the pollutant, thus it is interesting to reduce the appearance and energy of the A modes and enhance the B modes. Also, the most unfavourable mechanisms present in the generator modes are the streamwise dome and the wall-normal cap, since they might contribute in the concentration of the pollutants in the city. On the other hand, regarding the breaker mechanisms, the most favourable mechanisms (regarding the pollutant concentration) are the spanwise arrowhead-like structures that might help the flow with the pollutants to escape from the urban region. Finally, the tunnel-shaped vortex created when the arch vortex is broken, could help redirect the pollutants retained between the buildings to the atmosphere by creating a recirculation area between the positive and negative values of the mechanisms in the B modes. Further studies should be carried out in this line using multi-phase flows, connecting the aforementioned flow structures with the particle dispersion mechanisms.

RV acknowledges the financial support of the G\"oran Gustafsson foundation. AMS and SH were funded by Contract No. RTI2018-102256-B-I00 of Ministerio de Ciencia, innovaci\'on y Universidades/FEDER. AC and SLC acknowledge the grant PID2020-114173RB-I00 funded by MCIN/AEI/10.13039/501100011033. The computations carried out in this study were made possible by resources provided by the Swedish National Infrastructure for Computing (SNIC).

\input{biblio}

\end{document}

%% file: biblio.tex
%